\documentclass[doublecol]{epl2}

\usepackage{amssymb}

\title{
Contact mechanics of and Reynolds flow through saddle points
}
\subtitle{On the coalescence of contact patches and the leakage rate through 
near-critical constrictions}
\shorttitle{Contact mechanics of saddle points} 

\author{Wolf B. Dapp\inst{1} 
\and Martin H. M\"user\inst{1,2}}
\shortauthor{W. B. Dapp \etal}

\institute{                    
 \inst{1} 
Supercomputing Centre J\"ulich, Institute for Advanced Simulation,
FZ J\"ulich, 52425 J\"ulich, Germany\\
\inst{2} 
Lehrstuhl f\"ur Materialsimulation, Universit\"at des Saarlandes,
66123 Saarbr\"ucken, Germany }


\abstract{
We study numerically local models for the mechanical contact between two 
solids with rough surfaces. 
When the solids softly touch either through adhesion or by a small normal 
load $L$, contact only forms at isolated patches and fluids can pass through 
the interface.
When the load surpasses a threshold value, $L_{\rm c}$, adjacent patches 
coalesce at a critical constriction, i.e., near points where the interfacial 
separation between the undeformed surfaces forms a saddle point. 
This process is continuous without adhesion and the interfacial separation 
near percolation
is fully defined by scaling factors and the sign of $L_{\rm c}-L$. 
The scaling factors
lead to a Reynolds flow resistance which diverges as 
$(L_{\rm c}-L)^{-\beta}$ with $\beta = 3.45$. 
Contact merging and destruction near saddle points becomes discontinuous 
when either short-range adhesion or specific short-range repulsion 
are added to the hard-wall repulsion.
These results imply that coalescence and break-up of contact patches can 
contribute to Coulomb friction and contact aging.
}

\begin{document}

\maketitle

\section{Introduction}
When two nominally rough solids are pressed against each other, their surfaces 
tend to touch microscopically only at isolated points~\cite{Bowden56,Persson01}.
Simple models assume that real contact between two rough surfaces can be
decomposed into single-asperity, Hertzian-like contacts.
However, both experiment~\cite{Dieterich79,Swingler09} and large-scale 
simulations~\cite{Hyun04,Pei05,Campana08alone,Campana11,Putignano12} 
have revealed that the majority fraction of contiguous contact patches 
appear to be fractal
and to result from many (formerly) single-asperity contacts. 
While the latter are well studied, less is known about the way in which they
merge~\cite{Johnson85,Krithivasan07,ScaraggiEtAl2011,Yastrebov14}, 
in particular when adhesion is present.
For example, it is unclear if adhesive contact-patch coalescence and break-up
happen discontinuously.
Contact-patch coalescence also plays a role for seals:
it has been argued that their leakage rate is determined to a significant 
degree from the topology of the last critical constriction~\cite{Lorenz09EPL}, 
i.e., the neighborhood of the point, which, upon increasing load or adhesion,
is the first point to interrupt a percolating non-contact path.

In this work, we study local models for the merging of contact patches, or,
in the context of seals, the contact mechanics of critical constrictions.
The local gap topography of a critical constriction is such that the 
interfacial separation of the undeformed surfaces is (close to) a saddle point.
More precisely, near the percolation point, the gap is very small. 
Parallel to a just-blocked fluid channel, the gap opens, while in the 
orthogonal, in-plane direction, the gap is closed, because the interfacial 
separation of the undeformed surfaces decreases in that direction. 
Thus, studying the contact mechanics of saddle points entails the analysis of
critical constrictions and that of contact patch merging.
Here, we investigate not only the contact mechanics of (near-) critical 
constrictions, including a scaling analysis of the gap topography near the
percolation point, but also address the pertaining Reynolds flow with and
without adhesion. 

To simulate the contact mechanics of an isolated saddle point, we reinvestigate
rather simple models~\cite{Johnson85,Krithivasan07,Yastrebov14} for the 
(combined) surface roughness and for the interactions between the surfaces.
Yet, in addition to the hard-wall repulsion commonly assumed infinitely 
short-ranged in continuum mechanics, we also study the effects of finite-range 
attraction~\cite{Maugis92} and repulsion~\cite{Muser14Beilstein} between the 
surfaces.
Our motivation for additional repulsion stems from recent simulations, in which 
appropriately designed cohesive-zone models qualitatively reproduced the 
surface interactions mediated by a strongly wetting fluid inducing an 
effective negative surface energy of finite range~\cite{Muser14Beilstein}. 
The length-scale of these additional surface interactions is nevertheless short 
enough for interfacial interactions not to act far away from a contact line, 
i.e., our Tabor parameters~\cite{Maugis92} are $\gg 1$. 

\section{Models and Methods}
We assume linear elasticity and the small-slope approximation so that roughness
can be mapped to a rigid substrate and the elastic compliance to a flat counter 
body. 
The effective contact modulus is used to define the unit of pressure,
i.e., $E^* = 1$.
Elasticity is treated with Green's function molecular dynamics 
(GFMD)~\cite{Campana06} and the continuum version of the stress-displacement 
relation in Fourier space, 
$\tilde{\sigma}({\bf q}) = qE^* \tilde{u}({\bf q})/2$,
where ${\bf q}$ is a wave vector and $q$ its magnitude. 
Simulations are run in a force-controlled fashion.
After the external load has changed by a small amount, all degrees of 
freedom are relaxed until convergence is attained. 

Three models for the (combined) height profiles are employed.
Each profile has roughness only at a single wavelength $\lambda$, a 
root-mean-square height fluctuation of $h_0$, and a root-mean-square gradient 
of $2\pi h_0/\lambda$, which should be small enough for the small-slope 
approximation to be valid. 
In other words, all three lattices yield identical angle-averaged
spectra at non-zero wave numbers. 
First, we consider a square lattice
\begin{equation}
\frac{h_{\rm sq}(x,y)}{h_0}  =   2+
\cos\left(\frac{2\pi x}{\lambda} \right)+
\cos\left(\frac{2\pi y}{\lambda} \right),
\end{equation}
for which the simulation cell dimensions along $x$ and $y$ direction are chosen 
to coincide with the wavelength $\lambda$ of the height undulation, that is
$L_x = L_y = \lambda$. 
The two remaining lattices are triangular and hexagonal:
\begin{eqnarray}
\frac{h_{\rm tl}(x,y)}{h_0} & = & \sqrt{\frac{2}{3}}
    \left\{   \frac{3}{2} + 2\cos\left(\frac{\sqrt{3}\pi x}{\lambda}\right)
    \cos\left(\frac{\pi y}{\lambda}   \right) \right.
\nonumber\\ & & 
    \left.+\cos\left(\frac{2\pi y}{\lambda}\right)  \right\}
    \mbox{\,\, (hexagonal lattice)}\\
\frac{h_{\rm hl}(x,y)}{h_0} & = &  
    \sqrt{\frac{27}{2}}-\frac{h_{\rm tl}(x,y)}{h_0} \mbox{ (triangular lattice)}
\end{eqnarray}
for which the dimensions of the periodically repeated simulation cells 
are $L_x = 4\lambda/\sqrt{3}$ and $L_y = 2\lambda$.
This shape is relatively close to a square so that discretization 
corrections~\cite{Dapp14TL} are almost isotropic in our GFMD calculations, 
which decomposes the surfaces into $2^n \times 2^n$ elements. 

In each model, the height offset is set to yield a minimum height of zero. 
Other key geometrical attributes are:
Maximum height ($\sqrt{27/2}~h_0$, $4~h_0$, $\sqrt{27/2}~h_0$), 
mean trough depth ($\sqrt{6}~h_0$, $2~h_0$, $\sqrt{2/3}~h_0$),
skewness  ($-\sqrt{2/3}$, 0, $\sqrt{2/3}$), and
excess kurtosis (-1/2, -3/4, -1/2), each time in the order
hexagonal, square, and triangular lattice.

Two surfaces interact with a hard-wall constraint, i.e., they are not allowed
to overlap. 
In addition we assume cohesive zone models for the finite-range interactions
between the surfaces.
It is $\gamma = -\gamma_0 \exp\{-g/\rho\}$ for attractive forces, 
where $g=g(x,y)$ is the local gap or interfacial separation.
Typical values for both $\gamma_0$ and $\rho$ are small numbers in the 
appropriate unit system, e.g., $\gamma_0 = 0.2~E^* h_0^2/\lambda$ and 
$\rho = 0.04~h_0$.
To give one possible realization: If $E^*$ were 1~GPa, $h_0 = 10$~nm, and 
$\lambda = 1$~$\mu$m, then $\gamma_0 = 20$~mN/m,
which reflects the order of magnitude for the surface energy
of flat, non-polar, and chemically passivated surfaces.
Finite-range repulsion is modeled with a different functional form than 
adhesion, namely with 
$\gamma(x,y) = -\gamma_0 \exp\{-g(x,y)^2/2\rho^2\}$.
The detailed rationale for these functional forms is given 
elsewhere~\cite{Muser14Beilstein}.
In brief:
the exponential adhesive model mimics the large Tabor number limit
quite efficiently while allowing one to determine contact area in an 
unambiguous manner. 
The Gaussian repulsive interactions can simulate the effect of a last layer, 
which needs to be squeezed out
before two surfaces touch. 

\begin{figure}
\onefigure[width=8.5cm]{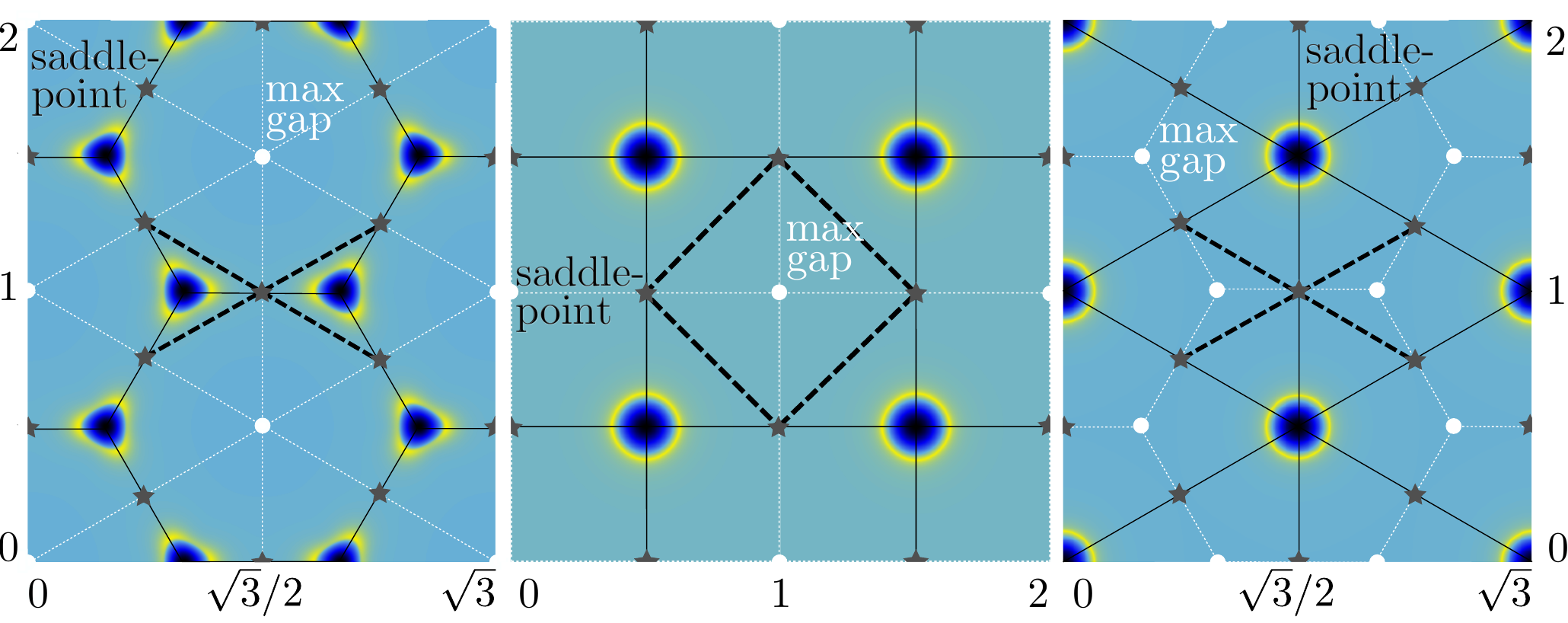}
\caption{\label{fig:forceFree}
Elastic displacements at zero load for the 
hexagonal, square, and triangular substrate lattice, from left to right.
Dark colors represent small displacements, that is, points of contact.
These are enveloped by yellow halos indicative of localized, adhesive necks. 
Thin black lines connect substrate height maxima, while thin white lines
connect substrate height minima or troughs, which are marked by white points.
Grey stars reflect saddle points. 
Only selected lines between nearest-neighbor saddle points are drawn
(thick dotted lines). 
}
\end{figure}
Figure~\ref{fig:forceFree} conveys an impression of our models.
It shows the elastic displacement for the investigated symmetries
in case of adhesion and zero external load.
The dark areas show the locations of small displacement and thereby
the structure of initial contact patches. 
The halos around contact points reflect bulges near the contact line
that are well known from single-asperity contacts with short-range adhesion.
One can also recognize that the displacements are rather constant near the 
height minima.
The latter lie on lattices dual to the height maxima for the considered cases. 
For example, the height minima (or gap maxima) form a triangular lattice
when the substrate heights lie on a hexagonal lattice, and vice versa. 
Saddle points form a kagome lattice in these two cases. 
For the square lattice, not only height maxima but also minima
and saddle points each form a square lattice. 

Flow through the interface is described by Reynolds thin-film equation,
which assumes a local conductivity that scales as the inverse third power
of the gap.
We use the \verb+hypre+ package~\cite{FalgoutEtAl2006} to solve the sparse linear system that
the discretized Reynolds equation can be expressed as. 
We employ the solvers supplied with \verb+hypre+ using the CG (conjugate 
gradient), or GMRES (generalized minimal residual) methods~\cite{SaadSchultz1986}, 
each preconditioned using the PFMG method, which is a  parallel
semicoarsening multigrid solver~\cite{AshbyFalgout1996}. 
Our in-house code is MPI-parallelized and uses HDF5 for I/O.

\section{Results}

\begin{figure}
\onefigure[width=8.0cm]{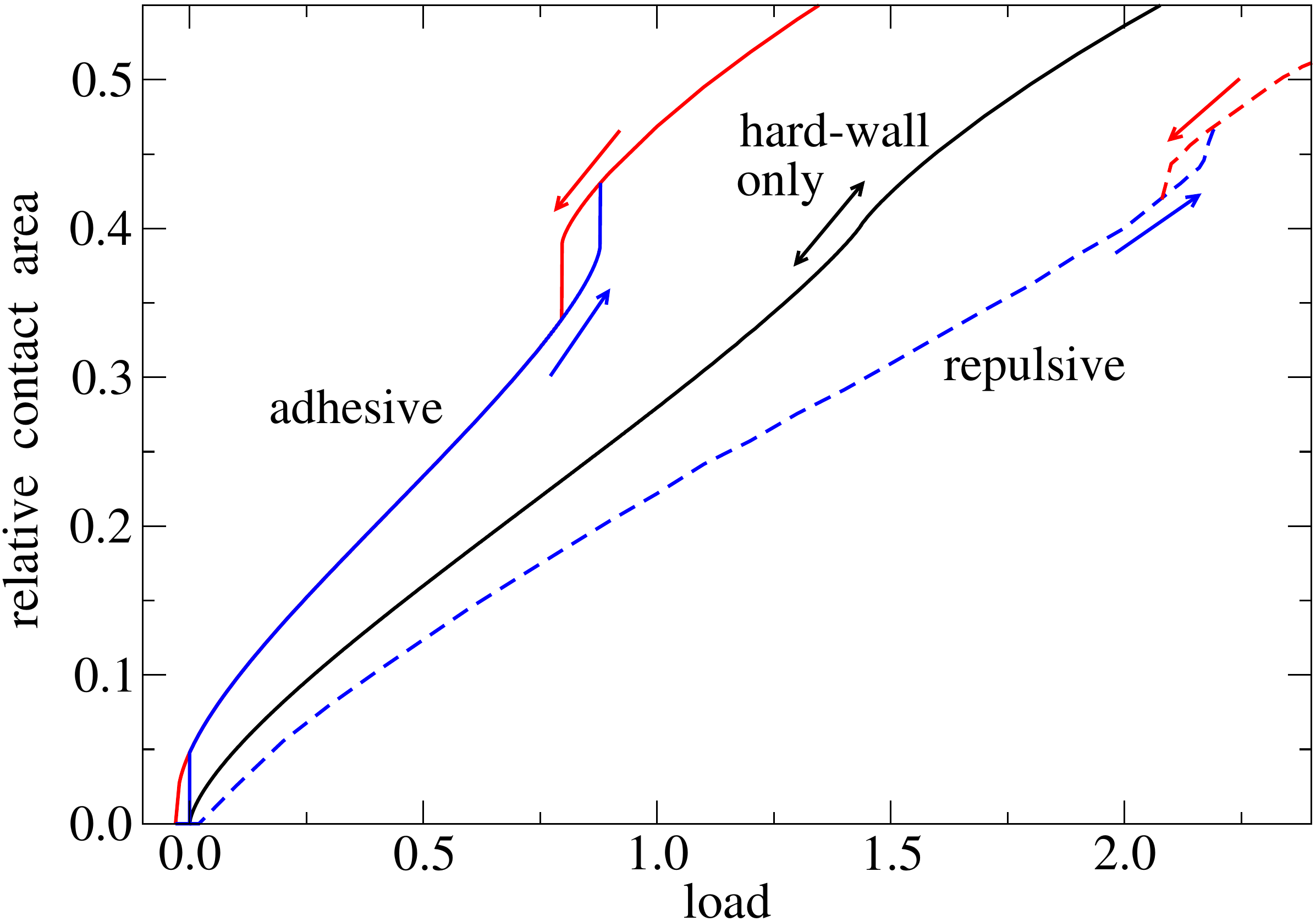}
\caption{\label{fig:sqLatAllAc}
Relative contact area as a function of load for the square-lattice substrate
and different finite-range interactions. 
Red (blue) lines refer to increasing (decreasing)  external load as marked
by arrows. 
Exponential adhesion (solid lines) and Gaussian 
repulsion (dashed lines) show the established hystereses 
at small contact area related to contact formation.
Unless solids interact solely by hard-wall repulsion, 
additional hystereses appear at relative contact areas near 0.4.
They reflect the contact-patch-coalescence and break-up instabilities. 
}
\end{figure}

A central question addressed in this study is whether the merging of two 
contact patches happens continuously or discontinuously.
Figure~\ref{fig:sqLatAllAc} confirms previous 
findings~\cite{Johnson85,Yastrebov14} that percolation is continuous 
when surfaces interact solely with hard-wall repulsion.
However, once finite-range repulsion (with the functional form defined in the 
method section) or short-range attraction are added, contact patches and
likewise fluid channels coalesce and break up through instabilities. 
The behavior is qualitatively similar for the two remaining models,
despite large quantitative differences.

\begin{figure}
\onefigure[width=8.0cm]{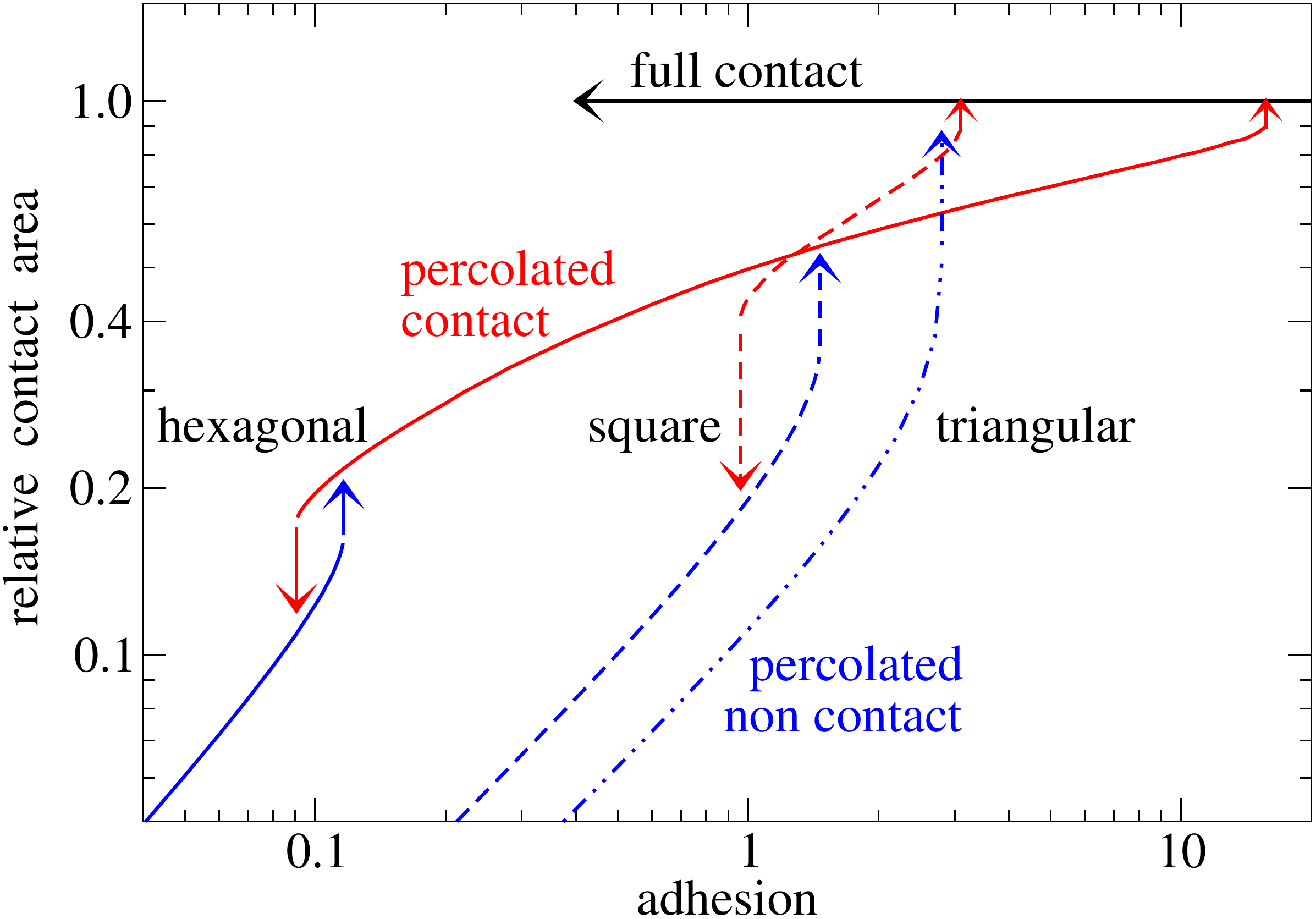}
\caption{\label{fig:adhAllLatt}
{Relative contact area} as a function of surface energy $\gamma_0$
for the three investigated models. 
The instabilities at lower values of $\gamma_0$ are related to percolation 
transitions. 
The single-wavelength, triangular substrate circumvents the percolated, partial
contact regime by transitioning directly from isolated contact patches to full 
contact.
}
\end{figure}

Contacts also coalesce discontinuously when driven entirely by adhesion.
This is revealed in Figure~\ref{fig:adhAllLatt}.
Despite similar conditions --- identical adhesion and surface spectra --- 
the quantitative differences between the three models are clearly borne out:
The hexagonal (triangular) lattice has by far the largest (smallest) 
propensity to form a coalesced or percolated contact area. 
This is because its peaks are rather blunt or obtuse (pointed or acute) 
and the ridges connecting the peaks have small (large) curvature.
However, the hexagonal (triangular) lattice has the smallest tendency
to go into full contact, because its troughs are rather deep (shallow)
and the curvature normal to the ridges are large (small) in magnitude. 
In the triangular lattice, the propensity to form full contact is even so
large that partial, coalesced contact is not even metastable at finite
adhesion and zero load. 
The square model is somewhere in between the two extreme cases.

While adhesion-free/load-driven and adhesion-driven/load-free percolation 
are qualitatively different in that the former is continuous and the latter is 
discontinuous, they happen at similar relative contact areas. 
The precise values in the purely load-driven case with only hard-wall repulsion
are $a_{\rm p} = 0.17826(11)$ (triangular lattice), $0.40185(6)$ 
(square lattice), and $0.67323(1)$ (hexagonal lattice).
Interestingly, the average of these numbers (0.418) is close to the percolation 
threshold identified for self-affine, isotropic, randomly-rough and
non-adhesive bodies (0.425)~\cite{Dapp12PRL}. 
In the adhesion-driven case, the percolation of contact is triggered at
relative contacts slightly below the given values of $a_{\rm p}$ and ends
at values slightly above $a_{\rm p}$. 
The reverse transitions from partial but coalesced to isolated contact patches 
also occur near the respective values for $a_{\rm p}$, though shifted to 
slightly smaller values. 
While only peripheral to this work, we briefly discuss the transition
to full contact.
In the purely force-driven case, full contact occurs above critical mean 
pressures of $p_{\rm fc} = \pi E^* \bar{t}/\lambda$, where $\bar{t}$ is the 
mean trough depth (stated in the model section).  
This relation is readily obtained from the single-wavelength, full-contact 
solution~\cite{Johnson85}.
(In the original work, an additional factor of $\sqrt{2}$ was included, 
as the wavelength was defined to be $\lambda/\sqrt{2}$.)
The purely adhesion-driven percolation roughly follows the force-driven 
percolation, e.g., the ratio $\gamma_{\rm fc}/\gamma_{\rm p}$ is largest
for the hexagonal (135) and smallest for the triangular (1) model. 
Pertinent ratios for the critical forces in the absence of adhesion
are $31$ (hexagonal) and $1.4$ (triangular). 
Here and in the following, we do not address the reverse transition from full 
to partial contact for adhesive surfaces, because it would correspond to
a Griffith-like fracture problem lacking an initial crack.
As such, for our and related adhesive laws, full contact becomes unstable at
tensile pressure $\lesssim \gamma_0/\rho$. 
The other adhesion-related results presented here barely depend on the precise 
choice of $\rho$ as long as   $\rho \ll h_0$.

\begin{figure}
\onefigure[width=8.5cm]{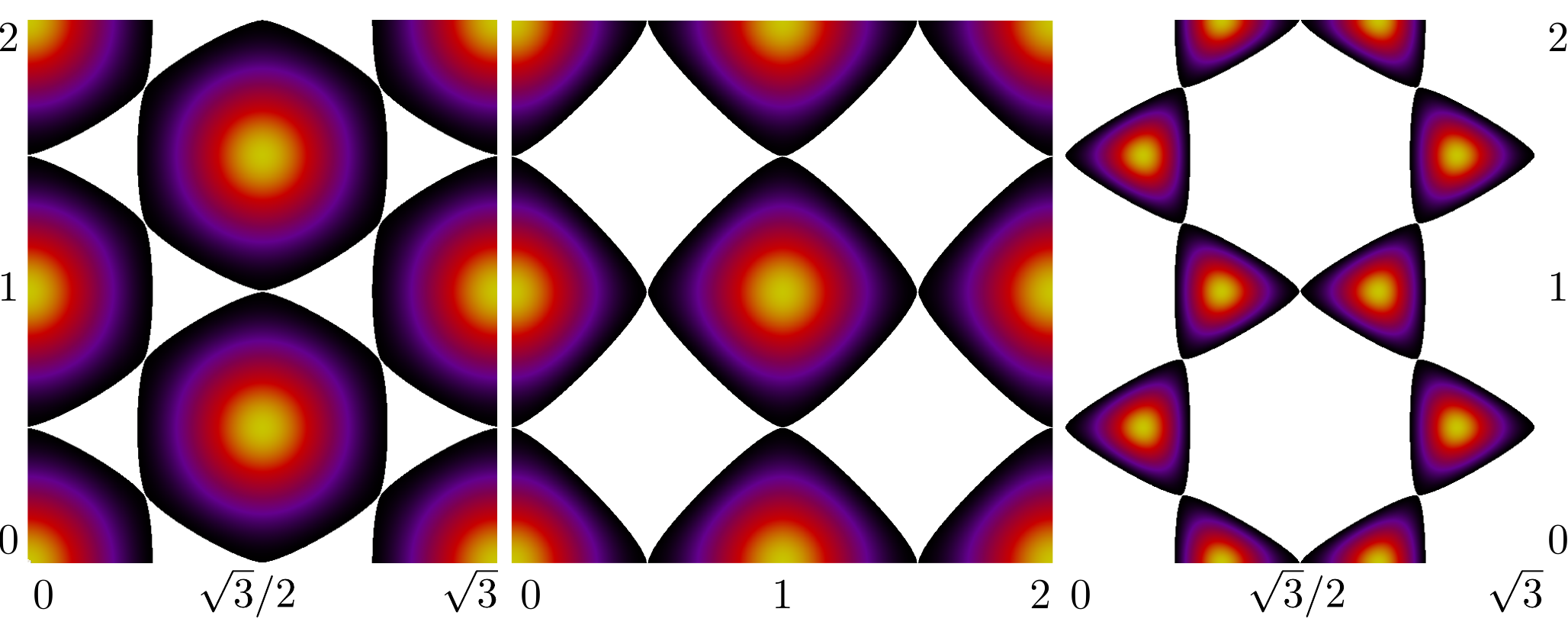}
\caption{\label{fig:percoGeometry}
Gaps (in color, where bright colors indicate large gaps and dark color small 
gaps) and contact (white) near the percolation threshold for the three 
investigated geometries. 
Systems are arranged in the same way as in Figure~\ref{fig:forceFree}. 
}
\end{figure}

The above results reveal a quite significant influence of local structure
on prefactors.
Some of the  differences can be rationalized from the contact and 
gap geometries near the percolation threshold, which are depicted in 
Figure~\ref{fig:percoGeometry} for simple hard-wall repulsion.
In coarse-grained representations the impression is conveyed that, 
depending on geometry, opening angles can range from acute to obtuse. 

Locally, however, all investigated saddle points are similar in that they sit at points 
of inversion symmetry, the curvatures of the substrate heights parallel to the
fluid-channel ($h''_{\vert\vert}$) are negative and those in orthogonal, 
in-plane direction ($h''_{\perp}$) are positive.
One can therefore expect that the solutions for gap and displacement are 
(locally) similar for the different symmetries, except for pre- and scaling 
factors.
In addition, percolation is continuous in the adhesion-free case, thereby
constituting a critical point. 
Thus, there should exist (quasi-) universal functions for the various fields in 
the vicinity of the critical point, e.g., for the gap, on which we focus 
exemplarily.
Following the ideas of the scaling hypothesis, the gap should then satisfy
\begin{equation}
g(l,x,y) = \vert l\vert^\zeta 
g^{\pm}\left( \frac{x}{\vert l\vert^\chi}, \frac{y}{\vert l\vert ^\upsilon} \right)
\label{eq:scaling}
\end{equation}
where $l \equiv (L-L_{\rm c})/L_{\rm c}$ is a reduced load, 
$g^{\pm}(\dots)$ denotes two master functions depending on the sign of $l$,
while $\chi$, $\upsilon$, and $\zeta$ are universal scaling exponents.
To superimpose the gap functions, not only for different values of $l$ but also
for different models --- or more generally speaking for different curvature 
ratios $\eta\equiv -h''_{\vert\vert}/h''_{\perp}$ characterizing the geometry 
of the saddle point --- appropriate unit systems for $x$, $y$, and $z$
need to be defined.
For example, for the square lattice, or $\eta = 1$, a reasonable choice is 
to define the minimum width and minimum height of the open channel as 1 
for $l = -0.01$  just like the length of the blocked channel for $l = 0.01$. 
For other geometries, or $\eta \ne 1$, the proper definition of units then
depends on $\eta$ and also on the sign of $l$. 

\begin{figure}
\onefigure[width=8.5cm]{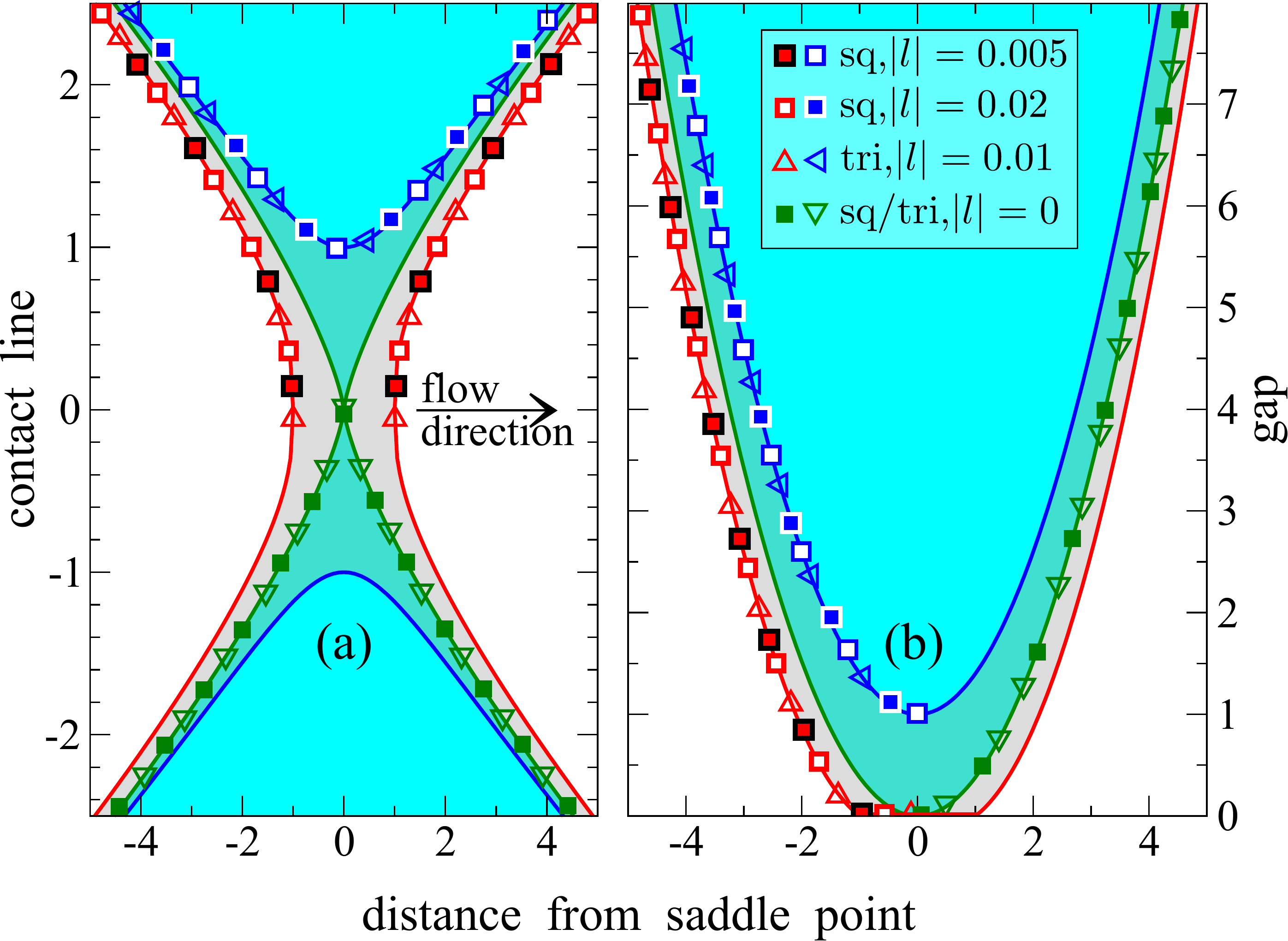}
\caption{\label{fig:GapGeometry} 
{\bf (a)} Contact line shape for normal loads below (blue), at (green), and, 
above (red) the critical load. 
{\bf (b)} Gap on the symmetry axis as a function of the distance from the
saddle point.
Different symbols represent different saddle-point geometries and different
colors different reduced loads $l$, as indicated in the caption. 
Solid areas and lines are obtained from analytical approximations to the 
contact shape. 
At a given value of $l$, distances are rescaled according
to equation~(\ref{eq:scaling}).
Units are chosen such that gap height, length, and width are 1 for 
$\vert l \vert = 0.01$ in the square model. 
}
\end{figure}

Numerically, we see equation~(\ref{eq:scaling}) to hold and demonstrate this
in Figure~\ref{fig:GapGeometry}.
Specifically, in panel (a), contact lines obtained for different reduced
loads and different geometries (exemplarily triangular and square) collapse 
onto their master curve that solely depends 
on the sign of $l$.
Thus, all (near-) critical constrictions from Figure~\ref{fig:percoGeometry},
locally adopt the contact line shapes from Figure~\ref{fig:GapGeometry} 
when viewed with sufficient magnification. 
Panel (b) shows that the gap can also be collapsed onto universal functions.

The numerical estimates for the exponents were obtained as follows:
We first identified $L_{\rm c}$ as accurately as possible.
We then ran simulations for 
$l = -0.002$, $-0.005$, $-0.01$, and $-0.02$. 
By analyzing how the gap at the saddle point scales with $l$, the
exponent $\zeta$ was determined to be $\zeta = 1.2\pm 0.01$.
The widths of the open channels were determined from the same set of 
simulations.
The contact line positions, $r_{\rm c}$, resulted from fitting the relation
$F(r_{\vert\vert} = 0,r_{\perp}) \propto \sqrt{r_\perp-r_{\rm c}}$ to the 
forces exerted by the substrate onto the elastic body near the contact line.
Here, the saddle point sat at $r_{\vert\vert} = 0$ and $r_{\perp} = 0$
in the local coordinate system.
This fitting procedure allowed us to determine the width of the open channel 
with sub-discretization resolution. 
All data was consistent with $\upsilon = 0.45 \pm 0.01$. 
To determine the length of blocked channels, simulations were
run for positive reduced loads and the same absolute values of $l$ as before. 
The pressure profile on the symmetry axis 
$(r_{\vert\vert}, r_{\perp} = 0)$ followed a Hertzian pressure profile 
very accurately, which allowed us to determine the length of the 
contact line on the symmetry axis to high precision and ultimately to 
ascertain that $\chi = 0.6 \pm 0.01$. 
Since the pressure profile on the contact ridges appears to be identical to 
that of a Hertzian contact, we believe that there exists an analytical solution,
in which case, we would expect the exponents to be simple rational numbers --- 
as in the Hertzian contact problem. 
We therefore speculate that $\chi = 3/5$, $\upsilon = 9/20$, and 
$\zeta = 6/5$ are the exact exponents.

\begin{figure}
\onefigure[width=8.5cm]{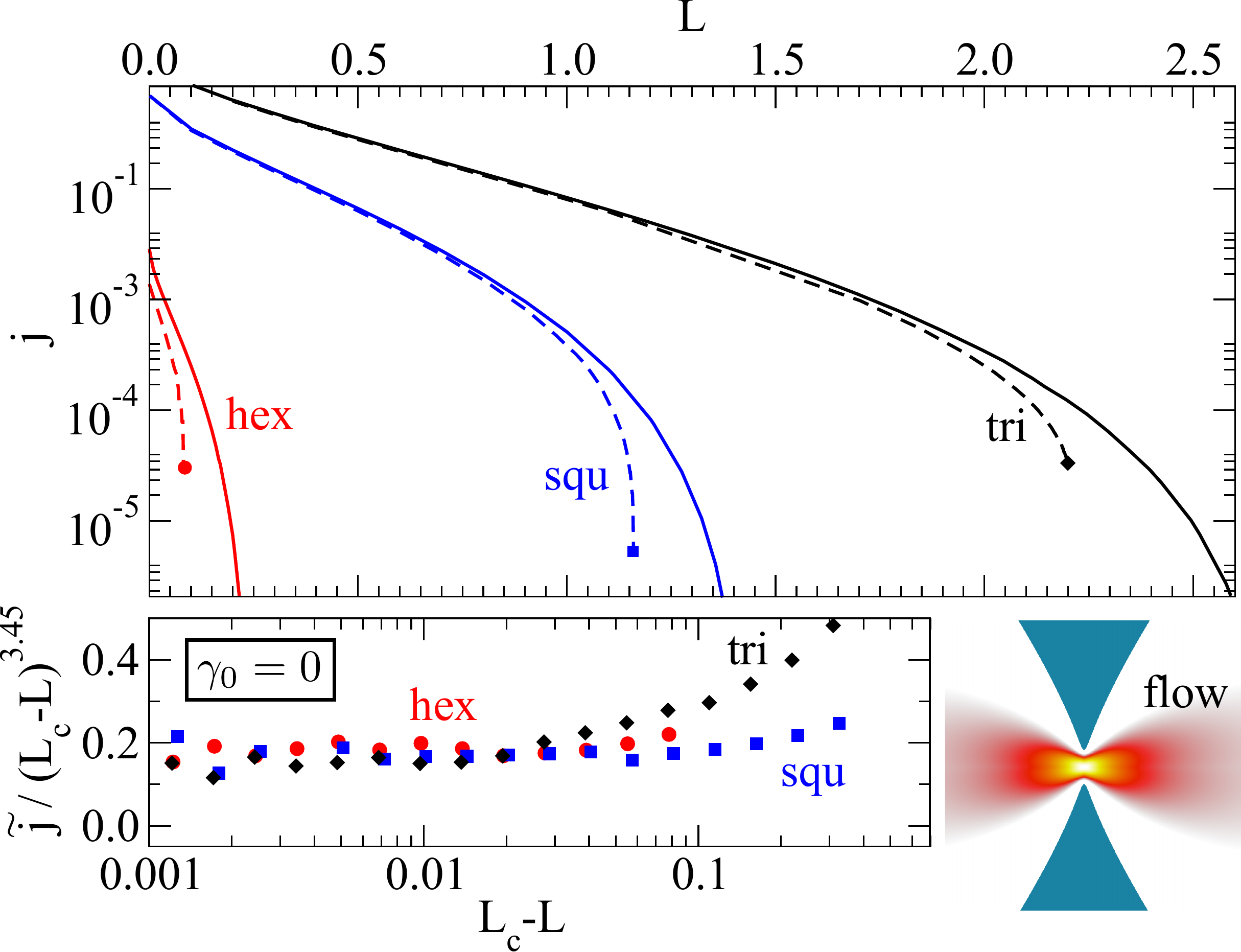}
\caption{\label{fig:FlowAsFunctionOfLoad} 
{\bf Top:} Fluid flow through the interface as a function of load.
Solid lines refer to $\gamma_0 = 0$ and dashed lines to $\gamma_0 = 0.05$.
The latter have end points, which are indicated by symbols.
{\bf Bottom left:} Normalized fluid flow as a function of $L_{\rm c}-L$, in the non-adhesive case.
{\bf Bottom right:} Visualization of the Reynolds thin-film
current density in the critical region. 
}
\end{figure}

Knowledge of the three exponents $\zeta$, $\chi$, and $\upsilon$ allows one 
to deduce the power law with which Reynolds flow is suppressed as the load
is increased toward $L_{\rm c}$.
Since the fluid-pressure gradient is non-negligible only near the critical
constriction, the resistance to fluid flow is proportional to the length of 
the channel.
Moreover, it is inversely proportional to the channel width and, according to the
Reynolds equation, the third power of the channel height. 
Thus, the resistance scales as
\begin{equation}
R(L) \propto (L_{\rm c}-L)^{-\beta} 
\end{equation}
with
\begin{equation}
\beta = 3\zeta+\upsilon-\chi,
\end{equation}
and a numerical value of $\beta = 3.45$. 
Figure~\ref{fig:FlowAsFunctionOfLoad} confirms these expectations for the 
investigated adhesion-free surfaces:
Near the percolation threshold, all three system disappear with the same power 
law, even if  prefactors may differ substantially. 
The exponent of $\beta = 3.45$ is reproduced quite accurately by the flow 
simulations. 
In the critical regime, the (scaled) current density always looks like the 
flow shown in the bottom right of Figure~\ref{fig:FlowAsFunctionOfLoad}.

The absolute flow through the contacts changes dramatically with the saddle 
point geometry even for fixed height spectra, in parts, of course, because the 
percolation point is so sensitive to the geometry. 
In other words, the leakage rate is strongly dependent on the curvature ratio 
$\eta$ or the skewness $s$ of the height distribution, potentially even more 
so than previously recognized~\cite{Lorenz10EPL}, see also~\cite{Yastrebov15}.
Specifically, the single-wavelength models have $\eta = (4,1,1/4)$ and
$s = (-\sqrt{2/3},0,\sqrt{2/3})$ for hexagonal, square, and triangular 
lattices, respectively. 
In addition, just as the contact area is discontinuous at the percolation 
transition, flow ceases and starts discontinuously with
the closing or opening of the constrictions once adhesion is considered. 

Our present analysis does not include plasticity, because it will mostly 
affect the points in earliest contact, namely the peak heights where the local 
pressures are highest.
We are mostly concerned with the areas with the most tenuous contact, 
and lowest contact pressures. 
Lateral plastic material flow from the peaks might slightly alter the detailed 
shape of the saddle point but such effects are beyond the scope of the present 
work.

The most problematic approximations for the continuation of the power law 
down to the nanoscopic vicinity of the final constriction are most likely 
that we neglect to consider (i) the existence of a slip length (and the associated 
deviation from a Poisseuille flow profile), (ii) confinement-induced viscosity 
change (via the orientation of molecules), as well as (iii) the non-smoothness of 
surfaces at nano-scales (where the continuum approximation breaks down).

However, we note that shear-thinning does not come into play for the flow 
analysis, not even very close to the percolation threshold. 
This is because the maximum shear rate of a Poisseuille flow
through a constriction is proportional to the gap height divided by the channel length.
As a consequence, the maximum shear rate disappears with $l^{\zeta-\chi}$ with
$\zeta-\chi \approx 0.6$, which 
we see confirmed by our simulation results. 

\section{Conclusions}

In this study we find that adjacent contact patches merge or break up 
discontinuously when finite-range interactions act in addition to  hard-wall 
repulsion.
This result has two main implications: 
First, in most cases, contact patches cannot be separated by a marginal distance
or joined via arbitrarily thin ridges, which
might explain why classical experiments~\cite{Dieterich79} visualizing the 
contact between two nominally flat solids as well as recent computer 
simulations including adhesion~\cite{Pastewka14} showed clearly separated 
contact patches, while large-scale simulations of {\it non-adhesive} 
solids~\cite{Hyun04,Pei05,Campana08alone,Campana11,Putignano12}
usually necessitate excessively high resolution to determine the connectedness
of contact patches.
Second, the contact-patch coalesence and break-up occur through first-order 
instabilities implying multistability. 
Thus, coalesence and break-up dynamics constitute a dissipation mechanism for 
Coulomb friction as any other instability induced by the relative sliding 
motion of two solids~\cite{Prandtl28,Muser02}.
Moreover, multistability entails thermal aging, which could be perceived as 
bulk relaxation and lead, for example, to a non-negligible 
time dependence of the leak rate of seals near percolation. 

In reference to Persson's contact mechanics~\cite{Persson01} and his 
single-junction leak-rate theory~\cite{Lorenz09EPL}, 
we also find two main implications.
First, we demonstrate that the contact mechanics of surfaces with single-wavelength
roughness and their leakage rate are highly sensitive to the way in which local maxima 
and minima are arranged.
For example, our triangular and hexagonal models seal at very different normal 
loads and relative contact areas, although the two geometries 
have identical height spectra. 
It might be possible to incorporate such effects
--- effectively representing a correlation of the phases for different
$\tilde{h}({\bf q})$ at fixed $\vert {\bf q} \vert$ ---
into the theory via appropriate generalizations.
Second, in the absence of adhesion, we find that the length and the width of a 
constriction are not similar in magnitude as assumed in Ref.~\cite{Lorenz09EPL}.
Instead they disappear with different power laws as  $L_{\rm c}$
is approached. 

When adhesion is absent, the contact mechanics of ``well-behaved'' saddle points
shows universal behavior near the percolation point. 
There, the gap is entirely determined by
the sign of the reduced load and scaling factors. 
Another aspect of the universal behavior is that the pressure profile on a 
newly-formed contact ridge along the transverse direction seems
identical to that of a Hertzian contact. 
We therefore see the possibility for closed-form analytical solutions of
the contact mechanics of a saddle point, even if formulating the boundary
conditions could be difficult.

Finally, we find for idealized conditions frequently assumed in continuum 
approaches (zero slip length, no confinement effects on viscosity, 
hard-wall repulsion only) that the Reynolds flow through a critical constriction 
follows $(L_{\rm c}-L)^\beta$ with $\beta = 3.45$. 
Since, for the given assumptions, most fluid pressure drops near a single
constriction in the immediate vicinity of the percolation 
point~\cite{Lorenz09EPL}, the leak rate of a seal that is
pressed against a randomly rough macroscopic solid would depend on load with 
the same power law as the isolated constriction. 
Despite this knowledge it is quite difficult to predict the 
{\it absolute} current because the detailed topography 
of the last constriction has a tremendous effect on prefactors but
is usually unknown. 
Still, the percolation exponent would result from local contact mechanics
and not, as commonly assumed in the theory of percolation~\cite{Stauffer91},
from the disorder at large length scales.
Whether the classical view is reestablished once adhesion is included remains 
to be seen. 

\acknowledgments
We thank the J\"ulich Supercomputing Centre for computing time.


\end{document}